\documentclass[FBSmath,ecsub]{FBSsuppl}
\usepackage{amsfonts}
\usepackage{amssymb}
\usepackage{epsfig}

\title{Axial Excitation of the $\Delta$ in Chiral Quark Models}

\author{S.~\v{S}irca\instnr{1,2},
        L.~Amoreira\instnr{3,4},
        M.~Fiolhais\instnr{5,4},
        B.~Golli\instnr{6,2}}
\instlist{Faculty of Mathematics and Physics,
          University of Ljubljana, Slovenia
     \and Jo\v{z}ef~Stefan Institute, Ljubljana, Slovenia
     \and Department of Physics,
          University of Beira Interior, Covilh\~a, Portugal
     \and Centre for Computational Physics, University of Coimbra, Portugal
     \and Department of Physics, University of Coimbra, Portugal
     \and Faculty of Education, University of Ljubljana, Slovenia}

\def\d{\mathrm{d}}
\def\half{{\textstyle{1\over2}}}

\def\th{{\scriptscriptstyle{3\over2}}}

\def\vec#1{\mbox{\boldmath$#1$}}
\def\svec#1{\mbox{{\scriptsize \boldmath$#1$}}}

\begin{document}

\maketitle
\begin{abstract}
We discuss electro-magnetic and weak axial N-$\Delta$ transition
amplitudes in the linear $\sigma$-model and the Cloudy-Bag Model
as typical representatives of chiral quark models.  We show that
good qualitative understanding of the transition can be obtained
in models which, in addition to the pion cloud, incorporate
a fluctuating $\sigma$-field inside the baryon.
\end{abstract}

\section{Introduction}

The present work was partly motivated by the experience gained
in the phenomenological description of the quadrupole
electro-excitation of the $\Delta$ within
the linear $\sigma$-model (LSM) and the Chromo-Dielectric Model (CDM)
in which the pionic degree of freedom was shown to play a dominant
role \cite{FGS}.  In these models, the pion cloud practically
saturates the electric and Coulomb quadrupole transition strengths
and qualitatively reproduces the $Q^2$-behaviour of the amplitudes.
Furthermore, the results for the ratios of electro-magnetic couplings
E2/M1 and C2/M1 at $Q^2=0$ were found to be rather insensitive
to the details of the models.  In the LSM, the absolute values
of the transverse helicity amplitudes were well reproduced,
although the underprediction of the strength at $Q^2=0$
remains an open question.  To pin down the pertinent model
ingredients, one therefore needs to probe non-zero values of $Q^2$
where effects of a possibly strong pion cloud in the interiors
of the nucleon and the $\Delta$ are manifested differently.

\smallskip

We anticipate that the theoretical investigation of axial N-$\Delta$
transition amplitudes \cite{llewellyn72} in chiral quark models
may reveal additional information on non-quark degrees of freedom in baryons.
Yet due to difficulties in consistent incorporation of the pion field,
the model predictions for these amplitudes are very scarce \cite{Mukh},
in contrast to the electro-magnetic sector.  Experimentally,
the structure of the weak axial transition currents
is explored by using weak probes \cite{neutrinos} or electron
scattering \cite{electrons}.   We have a stable world-average
for the dominant coupling $C_5^\mathrm{A}(0)$ \cite{mukh98},
but a very poor knowledge of $C_3^\mathrm{A}(0)$, $C_4^\mathrm{A}(0)$,
and the corresponding form-factors.  We calculated the axial amplitudes
in the Cloudy-Bag Model (CBM) and in the LSM.

\section{Calculation of the amplitudes}

In the linear $\sigma$-model and related classes
of models involving quarks interacting with chiral fields $\sigma$
and $\roarrow{\pi}$ the Hamiltonian can be written as
$$
H = H^0_q + H_\sigma + \int\d \vec{r}\left\{\half\left[\roarrow{P}_\pi^2
  + (\nabla^2 + m_\pi^2)\roarrow{\pi}^2\right]
  + U(\sigma,\roarrow{\pi}) + \roarrow{j}\roarrow{\pi}\right\}\>{,}
$$
where $j_a$ is the quark source, $\roarrow{P}_\pi$ is the pion
conjugate momentum, $H^0_q$ and $H_\sigma$ are the free-quark
and the $\sigma$-meson terms, and $U(\sigma,\roarrow{\pi})$
is the meson self-interaction term.  In the Cloudy-Bag Model
the $\sigma$-field and the $U$-term are absent,
while in the linear $\sigma$-model all terms are present and $U$
is the well-known
Mexican-hat potential.  The CDM has an additional scalar-isoscalar
field which mimics the glueballs of QCD and dynamically confines
the quarks \cite{CDM}.

\smallskip

Evaluating the commutator $[H,\roarrow{P}_\pi]$ between eigenstates
of the Hamiltonian $|\mathrm{N}\rangle$ and  $|\Delta\rangle$ 
(regardless of the model) we obtain a virial constraint of the form
\begin{equation}
 (-\Delta+m_\pi^2-(E_\Delta - E_\mathrm{N})^2)\langle\Delta\,|\,
  \pi_0(\vec{r})\,|\,\mathrm{N}\rangle   = 
  - \langle\Delta\,|\,J_0(\vec{r})\,|\,\mathrm{N}\rangle
\label{offdiagGT}
\end{equation}
where the source on the RHS of consists of the quark term and
the term originating from the meson self-interaction (if present):
$$
  J_0(\vec{r}) = j_0(\vec{r}) +
      {\partial U(\sigma,\roarrow{\pi})\over\partial\pi_0(\vec{r})}\>{.}
$$
In the CBM we assume the usual perturbative form for the pion
profiles using the experimental masses for the nucleon and $\Delta$,
which fulfills (\ref{offdiagGT}).  The method we used to impose this
constraint in the LSM is described in refs.~\cite{plbdraft,miniBled}.

\smallskip

In computing the transverse ($\tilde{A}^\mathrm{A}$),
longitudinal ($\tilde{L}^\mathrm{A}$),
and scalar ($\tilde{S}^\mathrm{A}$) transition helicity amplitudes
between states with definite four-momenta,
we interpret our localised model states as wave packets
of states with good linear momentum.  Extending
the method explained in \cite{Hemmert} we find, to order $k^2/M^2$:
\begin{eqnarray*}
C_6^\mathrm{A} & \!\!\!=\!\!\! & {M_\mathrm{N}^2\over k^2}\,
    \left[-\tilde{A}^\mathrm{A}_\th
  + \sqrt{3\over2}\tilde{L}^\mathrm{A}\right]\,
    {2M_\Delta\over M_\Delta+M_\mathrm{N}} \>{,}\\
C_5^\mathrm{A}  & \!\!\!=\!\!\! &
  -\sqrt{3\over2}\left(\tilde{L}^\mathrm{A}
  - {k_0\over k}\,\tilde{S}^\mathrm{A}\right)\,
     {2M_\Delta\over M_\Delta+M_\mathrm{N}}
  - {k_0^2-k^2\over M_\mathrm{N}^2}\,C_6^\mathrm{A} \>{,}\\
C_4^\mathrm{A}  &\!\!\! =\!\!\! &
 {M_\mathrm{N}^2\over kM_\Delta}\left[-\sqrt{3\over2}\,\tilde{S}^\mathrm{A}
 + {k_0k\over M_\mathrm{N}^2}\,{M_\Delta+M_\mathrm{N}\over2M_\Delta}\,
   C_6^\mathrm{A}
   \right] - {M_\mathrm{N}^2\over2M_\Delta^2}\,C_5^A \>{.}
\end{eqnarray*}

\smallskip

\noindent For a finite pion mass, the divergence of the axial
transition current is given by the PCAC relation
$
 \langle \Delta^+(p')\,|\,\partial^\alpha A_{\alpha\,a}\,|\,
    \mathrm{N}^+(p)\rangle
  = -m_\pi^2\,f_\pi\langle \Delta^+(p')\,|\,\pi_a(0)\,|\,
    \mathrm{N}^+(p)\rangle\>{,}
$
where $a$ is the isospin index and the transition matrix element
of the pion field with $a=0$ is related to the strong form factor by 
\begin{equation}
  \langle \Delta^+(p')\,|\,\pi_0(0)\,|\,\mathrm{N}^+(p)\rangle
 = \mathrm{i}{G_{\pi\Delta N}(Q^2)\over 2M_\mathrm{N}}\,
 {\bar{u}_{\Delta\mu}\,q^\mu u_\mathrm{N}\over Q^2 + m_\pi^2}\,
 \sqrt{2\over3}\>{.}
\label{piDN}
\end{equation}
Assuming that the pion pole dominates the $C^\mathrm{A}_6(Q^2)$
amplitude for $Q^2\rightarrow -m_\pi^2$, the resulting off-diagonal
Goldberger-Treiman relation \cite{llewellyn72,Hemmert,schreiner}
offers an alternative method to compute $C^\mathrm{A}_5(Q^2)$
from the strong $G_{\pi\mathrm{N}\Delta}$ form-factor,
$$
  C^\mathrm{A}_5(Q^2) =
  f_\pi\,{G_{\pi\mathrm{N}\Delta}(Q^2)\over2M_\mathrm{N}}\,
  \sqrt{2\over3} =
  {2M_\Delta\over M_\Delta + M_N}\sqrt{2\over 3}
  {f_\pi\over\mathrm{i}k}\int\d\vec{r}\, e^{\mathrm{i}\svec{k}\svec{r}}
   \langle\Delta||J_0(\vec{r})||\mathrm{N}\rangle\>{.}
$$

\section{Results and discussion}

The calculated $C_5^\mathrm{A}(0)$ is $25\,\%$ higher than
the experimental average, but the $Q^2$-dependence is reproduced
to within a few percent in terms of the dipole cut-off parameter.
A better result at $Q^2=0$ can be obtained by determining
$C^\mathrm{A}_5(Q^2)$ from the calculated strong
$\pi\mathrm{N}\Delta$ form-factor through the off-diagonal
Goldberger-Treiman relation, yet the $Q^2$-dependence becomes
steeper, with a cut-off of $\approx 0.80\,\mathrm{GeV}$.
The disagreement between the two approaches can be attributed
to an overestimate of the meson strength, a characteristic 
feature of LSM where only the meson fields bind the quarks.
(Still, the effect of the meson self-interaction is relatively
weak in the strong coupling constants.)  Essentially the same
trend is observed in the nucleon case where we obtain
$g_\mathrm{A}=1.41$.  The discrepancy with respect to
the experimental value of $1.27$ is commensurate with
the disagreement in $C_5^\mathrm{A}(0)$.  Unfortunately,
the overestimate of $g_\mathrm{A}$ and $G_\mathrm{A}(Q^2)$
in the LSM seems to persist even if the spurious centre-of-mass
motion of the nucleon is removed \cite{fizika383}.  

\smallskip

In the CBM the picture is reversed.  As it has been shown in
ref.~\cite{plbdraft}, only the quarks
contribute to the $C^\mathrm{A}_4$ and $C^\mathrm{A}_5$ amplitudes,
while $C^\mathrm{A}_6$ is almost completely dominated by the pion pole.
Hence the calculated values of $C^\mathrm{A}_5(0)$ are too small,
ranging from $68\,\%$ of the experimental estimate at $R=0.7\,\mathrm{fm}$
to only $56\,\%$ at $R=1.3\,\mathrm{fm}$.  The behaviour
of $C^\mathrm{A}_5(Q^2)$ is similar as in the pure MIT Bag Model
(to within $10\,\%$), with a fitted dipole cut-off
of $\sim 1.2\,\mathrm{GeV\,fm}/R$.
The off-diagonal Goldberger-Treiman relation is satisfied
in the CBM, but $C^\mathrm{A}_5(Q^2)$ calculated
from $G_{\pi\mathrm{N}\Delta}(Q^2)$
has a steeper $Q^2$-dependence with a cut-off of
$\sim 0.8\,\mathrm{GeV\,fm}/R$.
The large discrepancy can be partly attributed to the
fact that the CBM predicts a too low value for 
$G_{\pi\mathrm{NN}}$, and consequently for $G_{\pi\mathrm{N}\Delta}$.

\smallskip

The determination of the $C^\mathrm{A}_4$ is less reliable
because the meson contribution to the scalar amplitude
is very sensitive to small variations of the profiles.
However, the experimental value is very uncertain as well.
Neglecting the non-pole contribution to $\tilde{S}^\mathrm{A}$ and
$C_6^\mathrm{A}$, the value of $C_4^\mathrm{A}$ is dominated
by the term $-(M_\mathrm{N}^2/2M_\Delta^2)\,C_5^\mathrm{A}$,
in agreement with the popular value of $C_4^\mathrm{A}(0)=-0.3$.

\smallskip

In accordance with our experience in the electro-magnetic sector,
we find that the quark contribution alone strongly underestimates
the $C_5^\mathrm{A}$ amplitude.  If only a linear coupling of pions
to quarks is added, the situation does not improve since in such
a case the pion contribution to $C_5^\mathrm{A}$ vanishes.
On the other hand, the inclusion of meson self-interaction which
allows for a substantial deviation of the $\sigma$-field from its
vacuum value inside the baryon considerably increases $C_5^\mathrm{A}$.
The LSM overestimates this contribution, but this is not the case
in other chiral models which allow for a non-zero fluctuation
of the $\sigma$-field; in a version of the NJL model with nonlocal
regulators \cite{BGR} the contribution of sea quarks to the nucleon
$g_\mathrm{A}$ is more than a factor of two weaker than the equivalent
contribution of chiral mesons in the LSM.

\smallskip

This work was supported by FCT (POCTI/FEDER), Lisbon, and by
The Ministry of Science and Education of Slovenia.

\end{document}